\documentclass[amssymb,amsmath,aps,pra,twocolumn]{revtex4-2}

\usepackage[english]{babel}
\usepackage{graphicx}
\usepackage{color}
\usepackage{hyperref}
\usepackage{bm}
\usepackage{amsmath}
\usepackage{amssymb}
\usepackage{natbib}

\DeclareGraphicsExtensions{.eps,.pdf,.png,.jpg}

\begin{document}

\author{Yury Bliokh}
\email{bliokh@technion.ac.il}
\affiliation{Physics Department, Technion---Israel Institute of Technology, Haifa 320003, Israel}

\title{Pushing and Pulling Ponderomotive Forces in Wavepackets and Beat Waves}
		
\begin{abstract}

We consider ponderomotive forces acting on small particles in propagating wave packets (pulses). Specifically, we analyze simple point particles as well as composite dipole and dumbbell particles in the fields of forward-propagating (parallel phase and group velocities) and backward-propagating (antiparallel phase and group velocities) wave packets. Depending on the characteristics of the wave packet, particles may be pushed away from the wave source or pulled toward it. We also examine particle dynamics in the field of a beat wave generated by two forward-propagating waves with slightly different frequencies. Such a beat wave can emulate a periodic sequence of either forward- or backward-propagating pulses. In particular, this provides a simple mechanism for realizing pulling forces as employed in optical and acoustic `tractor beams'. 

\end{abstract}
	
\maketitle
	
\section{Introduction}
	
Ponderomotive forces determine the time-averaged motion of a particle in a fast-oscillating field of arbitrary nature. This concept is usually associated with the dynamics of particles in wave fields, but the first example is provided by a mechanical phenomenon: stabilization of the inverted pendulum by high-frequency vertical vibrations of its pivot: the so-called Kapitza pendulum \cite{Kapitsa1951}. Later, gradient ponderomotive forces have been employed for trapping of charged particles in an inhomogeneous oscillating electromagnetic field \cite{Gaponov-1958}. 

Nowadays, ponderomotive forces on particles in inhomogeneous wave fields of different natures are widely explored in various scientific domains. This includes laser-driven plasma-based accelerators of charged particles, as well as optical and acoustic manipulation of neutral particles ranging from individual atoms to macroscopic biological samples, see reviews \cite{Ashkin-2006, Grier-2003, Eseray-2009, Ozcelik-2018, Dholakia-2020, Toftul-2024}. 

Most optical and acoustic studies on wave-induced forces imply monochromatic fields (such as focused laser beams or standing ultrasonic waves) with inhomogeneous but stationary intensity distributions. At the same time, propagating wave packets (pulses) provide another fundamental class of wave fields, where spatial and temporal inhomogeneities act coherently. Surprisingly, the peculiarities of ponderomotive forces and particle dynamics in propagating wave packets have not been systematically analyzed (apart from the specific problem of wake-field acceleration of charged particles in plasma \cite{Tajima-1979, Eseray-2009}).

In this work, we examine the generic problem of 1D small-particle dynamics in propagating wavepacket fields. We explore the cases of point and composite particles, as well as wave packets with parallel and antiparallel phase and group velocities. Our study focuses on the pushing or pulling action of the ponderomotive force with respect to the wave source. Time-varying propagating fields have been shown as one of promising mechanisms for the `tractor-beam' pulling action \cite{Ruffner-2012, Mitri-2015, Lepeshov-2020}.

\section{Simple point particles} 

We consider a point particle in the 1D field of a quasi-monochromatic wave packet with a slowly varying amplitude: $a(x,t)\sin(\omega t-k x)$, where $k$ is the central wavenumber and $\omega$ is the central wave frequency. 
Introducing dimensionless coordinate of the particle, $\xi=|k|x$, and dimensionless time $\tau=\omega t$, the non-relativistic equation of motion of the particle can be written as    
\begin{equation}
\label{eq2}
\frac{d^2\xi}{d\tau^2}=\mathcal{F}(\xi,\tau)=a(\xi,\tau)\sin(\tau-s\xi).
\end{equation}
Here $\mathcal{F}$ is the force, $s={\rm sgn}(k)$, and we omit inessential constant factors. For a propagating wave packet, $a(\xi,\tau)=a(\xi-\eta_g\tau)$, where $\eta_g$ is the dimensionless group velocity of the wave. In the follows, we assume that $\eta_g>0$. In turn, the wave number $k$ can be either positive or negative, i.e., the wave phase velocity can be parallel ($s=1$) or antiparallel ($s=-1$) to the group velocity. Waves with parallel and antiparallel phase and group velocities will be referred to as the \textit{forward} and \textit{backward} waves, respectively.

{Let the wave amplitude be small and slowly (adiabatically) varying function:
\begin{equation}\label{condition1}
a\ll 1,\hspace{2mm}|da/d\xi|\ll1,\hspace{2mm} |da/d\tau|\ll1.
\end{equation} }
Then, using the perturbation approach, the solution of Eq.~(\ref{eq2}) can be presented as the sum of fast-oscillating, $\tilde{\xi}(\tau)$, and slow-varying, $\bar{\xi}(\tau)$, functions: $\xi(\tau)=\tilde{\xi}(\tau)+\bar{\xi}(\tau)$. In the first-order approximation in $a$ and $da/d\tau$, Eq.~(\ref{eq2}) yields the oscillating component: 
\begin{equation}
\label{eq3}
\tilde{\xi}=-a\sin(\tau-s\bar{\xi})-2\frac{\partial a}{\partial \tau}\cos(\tau-s\bar{\xi}).
\end{equation}
Substituting $\xi(\tau)=\tilde{\xi}(\tau)+\bar{\xi}(\tau)$ in Eq.~(\ref{eq2}) and expanding the right-hand side into the Taylor series, with subsequent averaging over the oscillation period, results in equation for the slow-varying component:
\begin{equation}
\label{eq4}
\frac{d^2\bar{\xi}}{d\tau^2}=-\frac{1}{4}\left(\frac{\partial a^2}{\partial\bar{\xi}}-2s\frac{\partial a^2}{\partial\tau}\right)\equiv \mathcal{F}_{\rm pond},     
\end{equation}
where $\mathcal{F}_{\rm pond}$ is the ponderomotive force acting on the particle. 

Expression (\ref{eq4}) shows that the ponderomotive force depends on the both spatial and temporal variations of the wave amplitude, where the corresponding contributions can be associated with the `gradient force' \cite{Gaponov-1958} and the `wave-momentum' force, respectively \cite{Bliokh-2022}. 
Since in a propagating wave packet, the spatial and temporal gradients of the amplitude are interconnected, these two forces act together.
Substituting $a(\xi,\tau)=a(\xi-\eta_g\tau)$ into Eq.~(\ref{eq4}), the ponderomotive force takes the form
\begin{equation}
\label{eq4a}
	\mathcal{F}_{\rm pond}=-\frac{1}{4}\left(1+2s\eta_g\right)\frac{da^2}{d\bar{\theta}},
\end{equation} 
where $\bar{\theta}=\bar{\xi}-\eta_g\tau$.

Hereafter, until Section~IV, we assume that the wave amplitude $a(\xi,\tau)$ has the form of an isolated propagating pulse, e.g., of a Gaussian form:
\begin{equation}\label{eq4c}
	a(\theta)=a_0\exp(-\theta^2/2\mathcal{L}_p^2),
\end{equation}
where {$$\mathcal{L}_p\gg1$$} is the dimensionless pulse length.  

The usual gradient force [the first term in Eq.~(\ref{eq4})] is directed oppositely to the amplitude gradient. This is also true for the wavepacket-induced force (\ref{eq4a}), when the group and phase velocities are co-directed, $s=1$. The situation changes if the group and phase velocities are antiparallel, $s=-1$. Depending on the value of the dimensionless group velocity $\eta_g$ (i.e., actually the ratio of the dimensional group velocity $v_g=d\omega/dk$ to the absolute value of the phase velocity $v_{\rm ph}=\omega/k$), the particle can be either repelled ($\eta_g<1/2$) or attracted ($\eta_g>1/2$) by the maximum-amplitude region. 

\begin{figure}[tbh]
	\centering {\includegraphics[width=\linewidth]{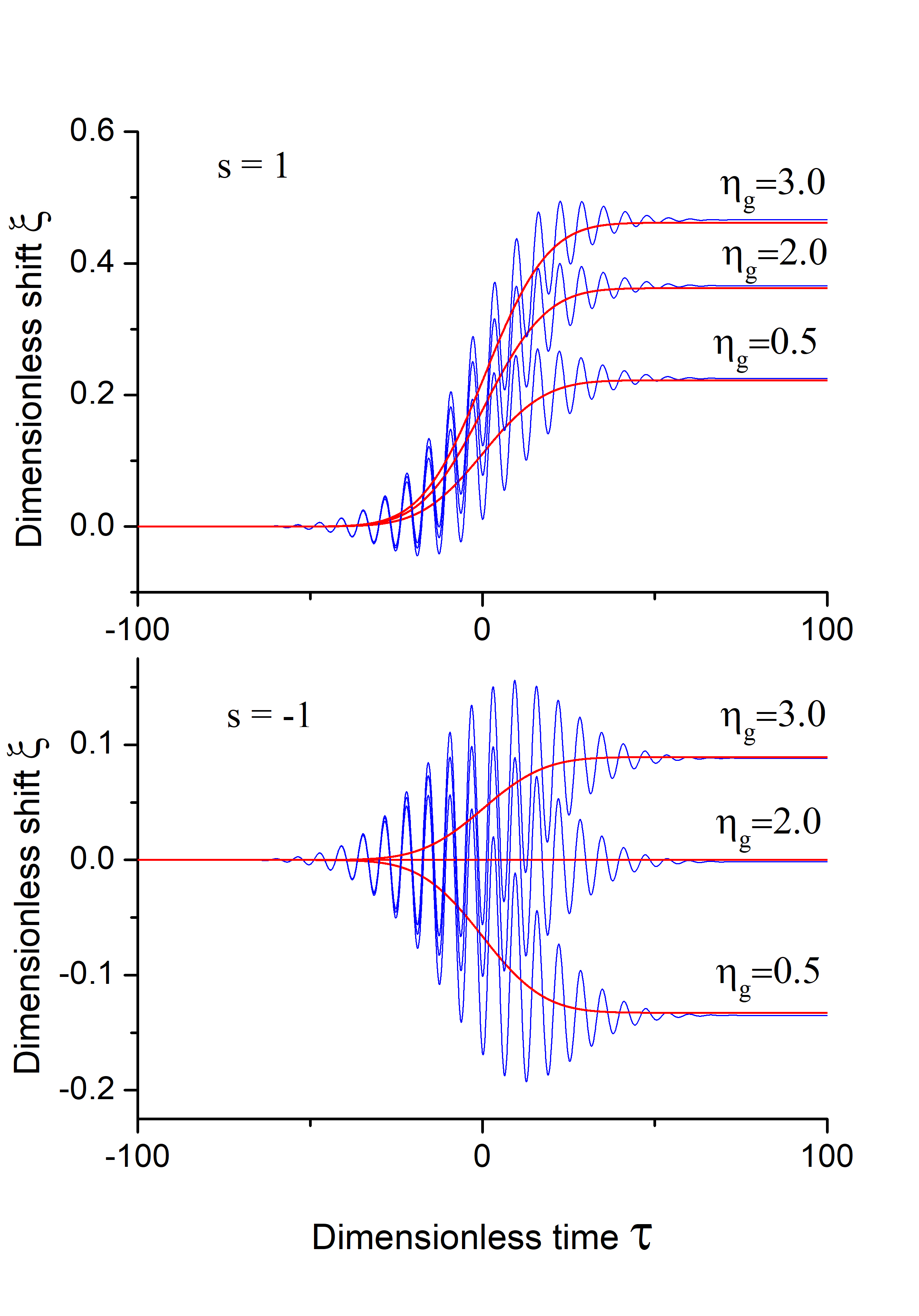}}
	\caption{Longitudinal shift of a simple point particle under the action of the pulses with different relations $\eta_g= |v_g/v_{\rm ph}|$ between the wave phase and group velocities. The blue oscillating curves are numerical solutions of Eq.~(\ref{eq2}), whereas the red smooth curves are solutions of Eq.~(\ref{eq4}). (a) The phase and group velocities are co-directed, $s=1$. (b) The phase and group velocities are anti-parallel, $s=-1$. 
        }
	\label{Fig1}
\end{figure}

Since the force (\ref{eq4a}) is proportional to the total derivative of the squared amplitude, it is conservative. Therefore, the particle, which was motionless before the pulse came, remains motionless after the pulse passes, but it changes its position. When the group and phase velocities are co-directed, the particle is always shifted along the pulse propagation direction. In contrast, a wave packet with antiparallel group and phase velocities shifts the particle forward or backward, or even leaves the particle position unchanged, depending on the value of the group velocity $\eta_g$. This result is confirmed by numerical solution of Eqs.~(\ref{eq2}) and (\ref{eq4}), and presented in  Fig.~\ref{Fig1}. 

\section{Composite particles}

We now consider simple models of composite particles consisting of two point particles connected by a massless rod. Namely, we examine three cases:

(1) Two particles of the same ``charge'' (i.e., response to the acting wave field) connected by a rigid rod, such that this {\it dumbbell} can move along the $x$-axis and change its orientation with respect to it. This can model particles of anisotropic shapes, such as spheroids \cite{Bruce-2021}.

(2) Similar composite particle but with opposite ``charges'' of two point particles. This model corresponds to a {\it permanent dipole}, such as molecules with an intrinsic electric-dipole moment \cite{Barry-2014}.

(3) Induced dipole particles, for which the dipole moment (the rod length) is proportional to the applied wave field. Such particles are ubiquitous for optical and acoustic systems \cite{Maher-2012, Drinkwater-2020}.   

\begin{figure}[tbh]
	  \centering {\includegraphics[width=0.8\linewidth]{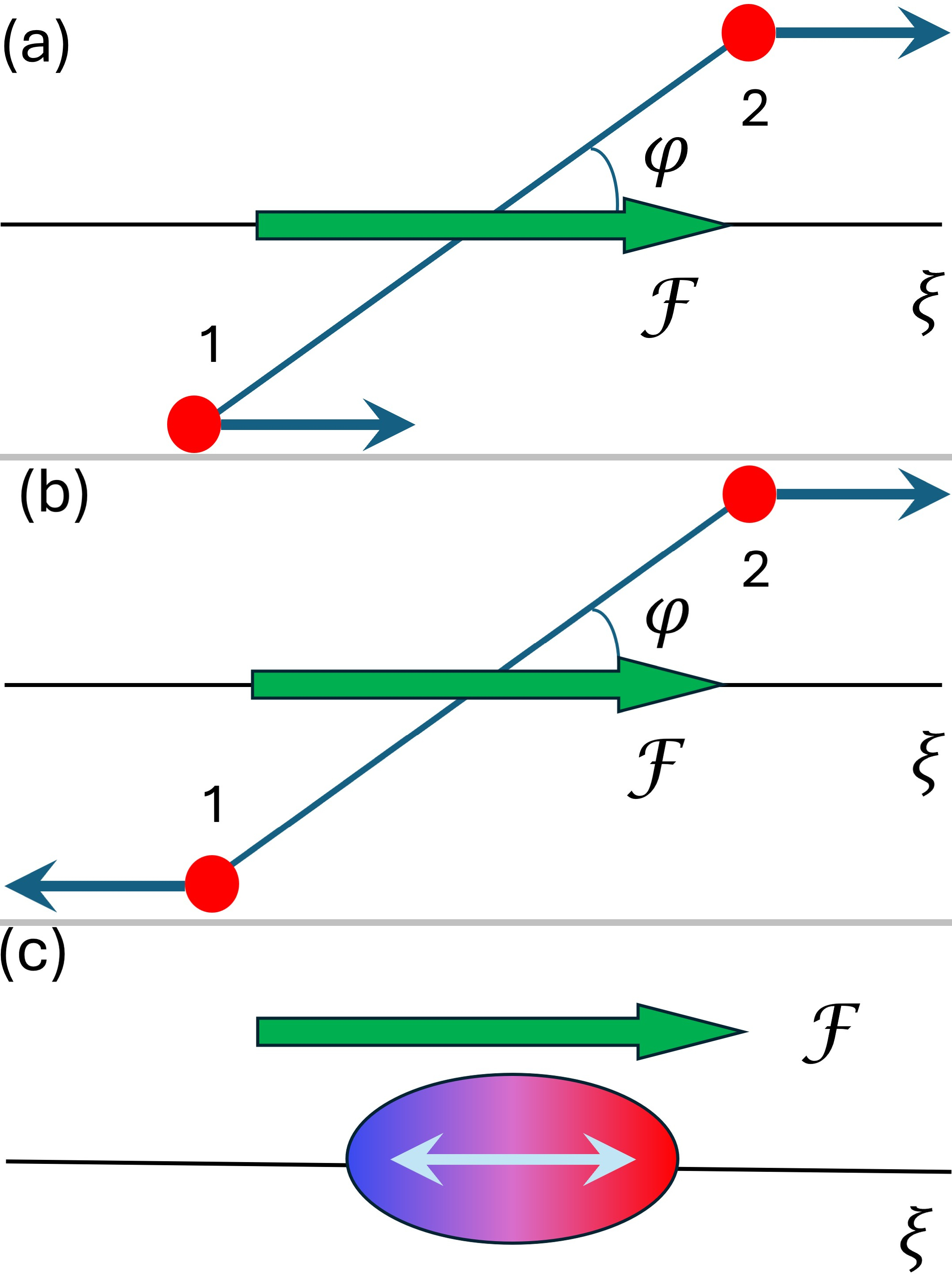}}
	\caption{Schematics of composite particles. (a) A dumbbell particle composed of two sub-particles with equal ``charges'' ($\sigma=1$) connected by a massless rod of length $d$. (b) A permanent-dipole particle composed of two sub-particles with opposite ``charges'' ($\sigma=-1$) connected by a massless rod. (c) An induced-dipole (polarizable) particle, whose dipole moment (the rod length) is proportional to the applied field $\mathcal{F}$.}
	\label{Composite}
\end{figure}

The composite particles in the cases (i) and (ii) possess an additional rotational degree of freedom, and their motion is described by the position of the center of mass, $\xi$, and the angle $\varphi$ between the rod and the $x$-axis, see Fig.~\ref{Composite}. 
This brings about the equations of motion for the translational and rotational degrees of freedom:
\begin{align}
	\frac{d^2\xi}{d\tau^2}&=\frac{1}{2}\!\left[\sigma \mathcal{F}\!\left(\xi-\frac{d}{2}\cos\varphi\right)+ \mathcal{F}\!\left(\xi+\frac{d}{2}\cos\varphi\right)\right]\!,\label{eq_cm}\\
	\frac{d^2\varphi}{d\tau^2}&=\frac{d}{2I}\sin\varphi\!\left[\sigma \mathcal{F}\!\left(\xi-\frac{d}{2}\cos\varphi\right)- \mathcal{F}\!\left(\xi+\frac{d}{2}\cos\varphi\right)\right]\!.\label{angle}
\end{align}
Here, $d$ is the dimensionless distance between the two point particles, $I=d^2/2$ is the corresponding moment of inertia of the composite particle, $\sigma=1$ for the dumbbell model (i), and $\sigma=-1$ for the permanent dipole model (ii). For the induced dipole case (iii), $\sigma=-1$, it is always aligned with the force, so that $\varphi \equiv 0$, and the distance between the point particle is proportional to the wave-induced force, $d= \alpha \mathcal{F}$ ($\alpha$ is the polarizability coefficient). 

{ Let the distance between point particles be small compared to the wavelength (the Rayleigh-particle limit):
\begin{equation}\label{condition2}
d\ll1.
\end{equation}.
Then Eqs.~\eqref{eq_cm} and \eqref{angle} can be simplified:}
\begin{align}
\frac{d^2\xi}{d\tau^2}&\simeq\frac{1}{2}\left[(1+\sigma)\mathcal{F}+\frac{d}{2}(1-\sigma)\cos\varphi \mathcal{F}^\prime\right]\!,
\label{cm2}\\
\frac{d^2\varphi}{d\tau^2}&\simeq\frac{1}{d}\sin\varphi\left[(\sigma-1) \mathcal{F}-\frac{d}{2}(1+\sigma)\cos\varphi \mathcal{F}^\prime\right]\!.
\label{angle2}
\end{align} 
where 
\begin{align}    
\label{Fprime}
\mathcal{F}^\prime \equiv \frac{d\mathcal{F}}{d\xi} &= \frac{da}{d\theta}\sin(\tau-s\xi)- sa(\theta)\cos(\tau-s\xi)\nonumber\\
&\simeq -sa(\theta)\cos(\tau-s\xi).
\end{align}
Equations~(\ref{cm2})--(\ref{Fprime}) contain only the leading terms. The role of the discarded terms $\propto d^2$ and $\propto da/d\theta$ will be discussed below. 

\subsection{Dumbbell}
\label{sec:Dumbbell}

\begin{figure}[tbh]
\centering {\includegraphics[width=\linewidth]{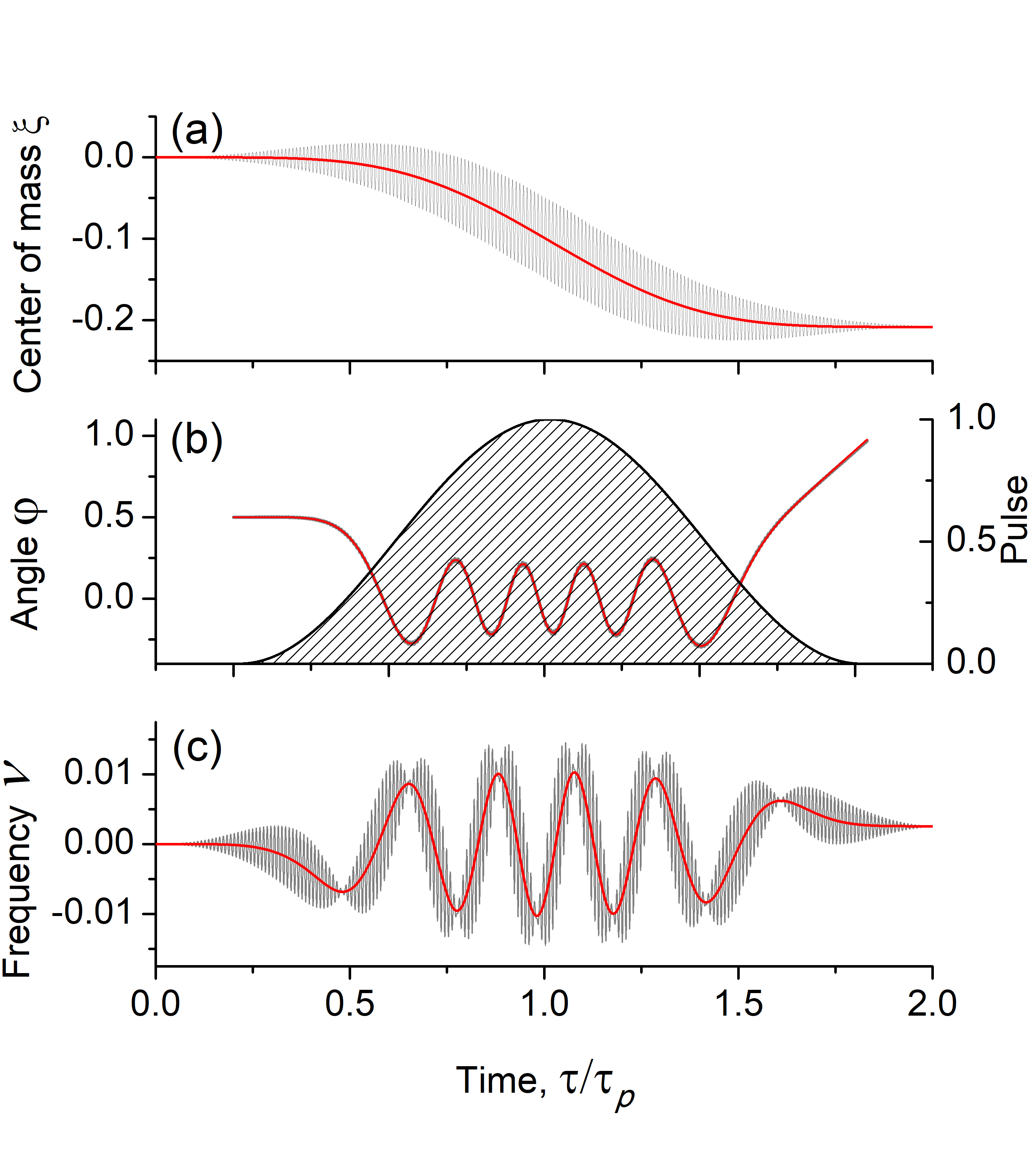}}
    \caption{Evolution of the center of mass (a), rotation angle $\varphi$ (b), and frequency $\nu=d\varphi/d\tau$ of a dumbbell particle under the action of a backward-wave ($s=-1$) pulse. Parameters of the pulse are: $a_0=0.05$, $\eta_g=0.75$. The dumbbell rod length is $d=0.1$. The gray lines are solutions of the complete Eqs.~(\ref{eq_cm}) and (\ref{angle}); the red lines are solutions of the time-averaged Eqs.~(\ref{eq4}) and (\ref{eq_phi}); the shaded area in (b) indicates the pulse shape. Time is normalized to the pulse duration $\tau_p=\theta_0/\eta_g$.}
\label{Dumbbell}
\end{figure}

We first consider the case of a dumbbell particle. 
The motion of the center of mass in the wave packet field is described by Eq.~(\ref{cm2}) with $\sigma=1$, which coincides with the equation of motion (\ref{eq2}) of a point particle and can be considered separately from Eq.~(\ref{angle2}). This results in the time-averaged evolution described by Eqs.~(\ref{eq4}) and (\ref{eq4a}). The time-averaged motion of the angle $\varphi$ can be obtained in a similar way:
\begin{equation}\label{eq_phi}
	\frac{d^2\bar{\varphi}}{d\tau^2}=-\frac{1}{4}a^2(\bar{\theta})\sin2\bar{\varphi}\left(1+\cos2\bar{\varphi}\right).
\end{equation} 
Examples of the numerical solutions of the approximate time-averaged Eqs.~(\ref{eq4}) and (\ref{eq_phi}) and exact Eqs.~(\ref{eq_cm}) and (\ref{angle}) are shown in Fig.~\ref{Dumbbell}. 

Variations of the argument $\bar{\theta} = \bar{\xi} - \eta_g\tau$ along the dumbbell trajectory are induced mainly by the time $\tau$ (the particle shift $\bar{\xi}$ is small). Therefore, the amplitude $a$ in Eq.~(\ref{eq_phi}) can be considered as a function of time only. Introducing a new time variable $\tau_*=\int_0^\tau d\tau^\prime a(\tau^\prime)$, which varies from 0 to some finite value $\tau_{*}^{\rm max}$, one can present Eq.~(\ref{eq_phi}) as follows:
\begin{equation}\label{eq_phi2}
	\frac{d^2\bar{\varphi}}{d\tau_*^2}+\frac{1}{a}\frac{da}{d\tau_*}\frac{d\bar{\varphi}}{d\tau_*}=-\frac{d\mathcal{U}}{d\bar{\varphi}},  
\end{equation}
where $\mathcal{U}=-1/4\,\cos^4\bar{\varphi}$ is the ponderomotive potential for the rotational degree of freedom. The second term on the left-hand side of Eq.~(\ref{eq_phi2}) can be neglected when the wave packet duration $\mathcal{T}=\mathcal{L}/\eta_g$ is large enough: 
{$$a_0\mathcal{T}\gg 1.$$} Then, Eq.~(\ref{eq_phi2}) becomes the equation of motion of a point particle in the potential well $\mathcal{U}$. 
Its solution describes {\it slow} oscillations of the averaged angle $\bar{\varphi}$. 
The substitution $\tau\rightarrow\tau_*$ deforms the time scale, but does not change the amplitude of these oscillations, which remains approximately constant inside the wave packet, as shown in Fig.~\ref{Dumbbell}(b). After the wave packet passes, the particle remains rotating with a constant angular velocity $d\bar{\varphi}/d\tau$.

Thus, the initially motionless dumbbell acquires rotational motion, and, hence, some energy from the wave packet. However, the wave packet cannot transfer energy without linear momentum. Therefore, the particle must also gain some linear momentum, and, unlike the point particle, the final dumbbell's velocity $d\xi/d\tau$ must differ from its initial velocity.  
This is inconsistent with the conservative character of the gradient force in Eqs.~(\ref{eq4}) and (\ref{eq4a}) for the translational motion. This controversy originates from the terms neglected in the transition from Eqs.~(\ref{eq_cm}), (\ref{angle}) to the simplified Eqs.~(\ref{cm2})--(\ref{Fprime}) and (\ref{eq4}), (\ref{eq4a}); it will be analysed in Section~\ref{sec:Transfer}.  

\begin{figure}[t]
	\centering {\includegraphics[width=\linewidth]{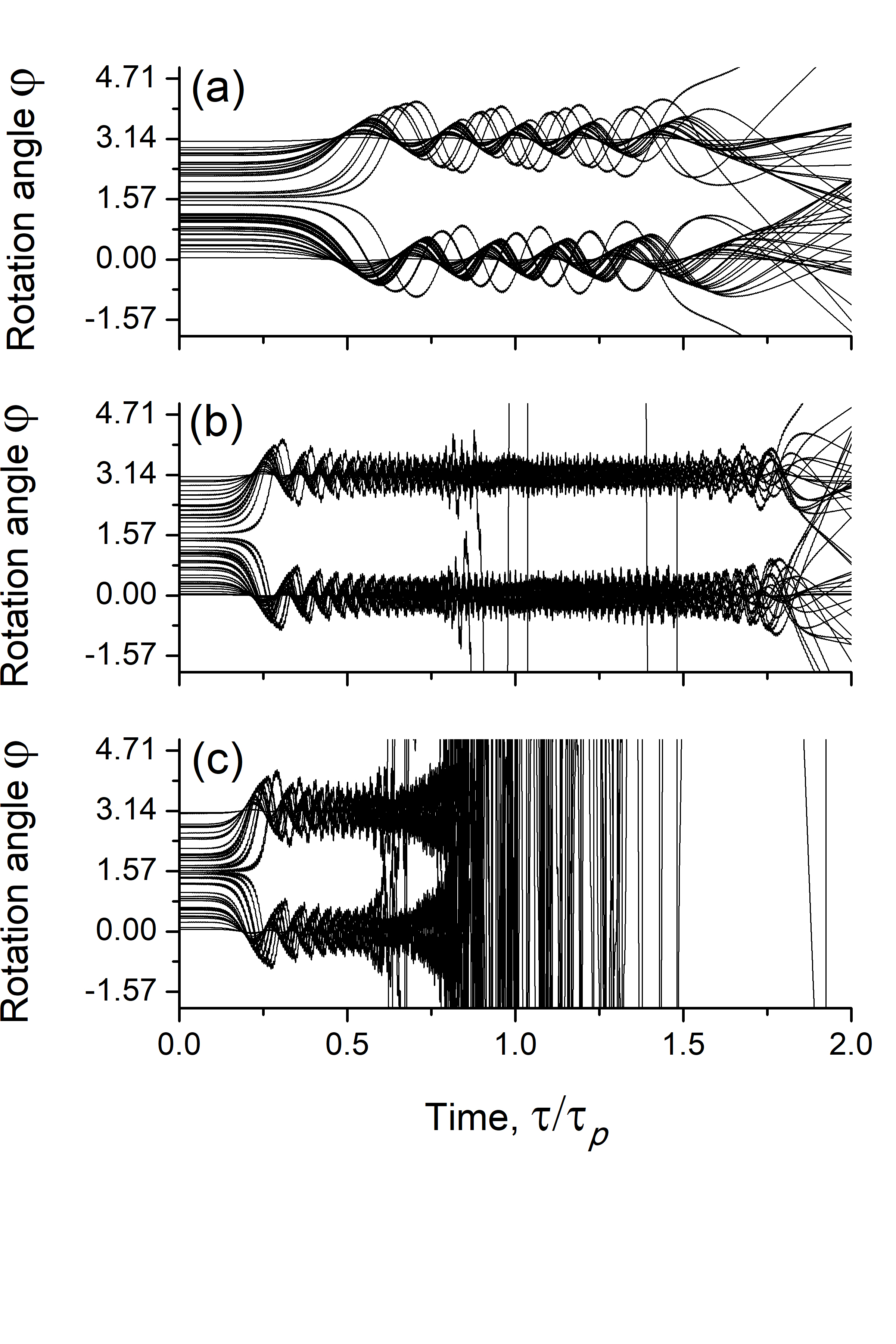}}
	\caption{Angular oscillations of 50 permanent-dipole particles, whose initial rotation angles $\varphi(0)$ are randomly distributed over the interval $(0,\pi)$. Time is normalized to the pulse duration $\tau_p=\theta_0/\eta_g$. The pulse amplitudes are: (a) $a_0=0.002$, (b) $a_0=0.02$, and (c) $a_0=0.03$. The frequency of oscillations increases as the wave amplitude increases, while the amplitude of oscillations remains practically unchanged.}
	\label{Dipole_New}
\end{figure}

\subsection{Permanent dipole}

Let us now consider the motion of a permanent dipole particle, case (ii), described by Eqs.~(\ref{cm2}) and (\ref{angle2}) with $\sigma=-1$. Since $d\ll 1$, the right-hand side of Eq.~(\ref{cm2}) is much less than the right-hand side of Eq.~(\ref{angle2}). Therefore, here we neglect the change in the center-of-mass position $\xi$, and consider only the dynamics of the rotation angle $\varphi$:
\begin{equation}\label{angle3}
\frac{d^2\varphi}{d\tau^2} =-\frac{2}{d}\sin\varphi \left. a(\theta)\sin(\tau-s\xi)\right|_{\xi=\rm const}.
\end{equation}
(The slow motion of the center of mass will be considered in the next subsection.) Representing the angle as the sum of fast-oscillating and slow-varying parts, $\varphi = \tilde \varphi + \bar \varphi$, we derive the time-averaged equation of motion:
\begin{equation}\label{angle4}
	\frac{d^2\bar{\varphi}}{d\tau^2}=-\frac{a^2}{d^2}\sin2\bar{\varphi}.
\end{equation}
Notably, this equation coincides with the equation of motion of the {\it Kapitza pendulum} in the absence of gravitation field \cite{Landau-1982}. Namely, it describes oscillations near the equilibrium positions $\bar\varphi=0$ and $\bar\varphi=\pi$ in the ponderomotive potential well $\mathcal{U}=a^2/d^2\sin^2\bar{\varphi}$. Therefore, the angular motion of a permanent dipole is qualitatively similar to the dumbbell case (i). 
The essential quantitative difference is that the condition Eq.~\eqref{condition1}, which allows the separation of the fast and slow motions, is now replaced by a stronger condition {$$a_0\ll d\ll 1,$$} which can be violated even when the pulse amplitude is small. 

Figure~\ref{Dipole_New} shows numerical solutions of exact Eqs.~(\ref{eq_cm}) and (\ref{angle}) for 50 dipoles, whose initial orientation angles $\varphi(0)$ are randomly distributed over the interval $(0,\pi)$. When $a_0/d \ll 1$, the angles oscillate near the equilibrium points $\varphi=0$ and $\varphi=\pi$ (Fig.~\ref{Dipole_New}a), in agreement with the approximate Eq.~(\ref{angle4}). As the small parameter $a_0/d$ increases, the oscillation frequency increases but the amplitude is not affected much. Furthermore, a fraction of the dipoles leave the equilibrium regime of finite angular oscillations, and starts to {\it rotate} rapidly, and can even occasionally change the rotation direction (Fig.~\ref{Dipole_New}b). The number of such rotating dipoles increases with the pulse amplitude, and for $a_0/d \sim 1$ the angular motion becomes chaotic (Fig.~\ref{Dipole_New}c), as described for the Kapitza pendulum \cite{McLaughlin-1981, Water-1991}.

\subsection{Energy and momentum transfer}
\label{sec:Transfer}

In deriving Eqs.~\eqref{cm2} and \eqref{angle2} the terms $\sim d^2\ll 1$ were omitted. As a result, the motion of dumbbell's center of mass was separated from the rotational motion. Taking account of these terms establishes a connection between the longitudinal and angular motions of the dumbbell and refines the equation of the center-of-mass motion:
\begin{equation}\label{dumbb-3}
\frac{d^2\xi}{d\tau^2}=\left(1-\frac{d^2}{8}\cos^2\varphi\right)\mathcal{F}.
\end{equation}
The evolution of the angle $\varphi$ is described by Eq.~\eqref{eq_phi}, which is a non-conservative equation of motion of a particle in a non-stationary potential well. The same is true for the dipole's angular motion,  Eq.~\eqref{angle4}.   

As we discussed in Section~\ref{sec:Dumbbell}, rotation of a composite particle after the wave packet passes implies that the wave energy and momentum are partially transferred to the particle.
{In contrast to the simple-particle case, this transfer is possible because of the excitation of an \textit{internal} degree of freedom: composite particle's rotation.}
From energy and momentum conservation laws, it follows that the ratio of the kinetic energy ($\mathcal{W}$) and 1D momentum ($\mathcal{P}$) of the particle must be equal to the ratio of the frequency and wavevector in the wave, i.e., in the dimensionless variables, $\mathcal{P}/\mathcal{W}=s=\pm 1$ \cite{Whitham-1999}. 
 
\begin{figure}[tbh]
	  \centering {\includegraphics[width=\linewidth]{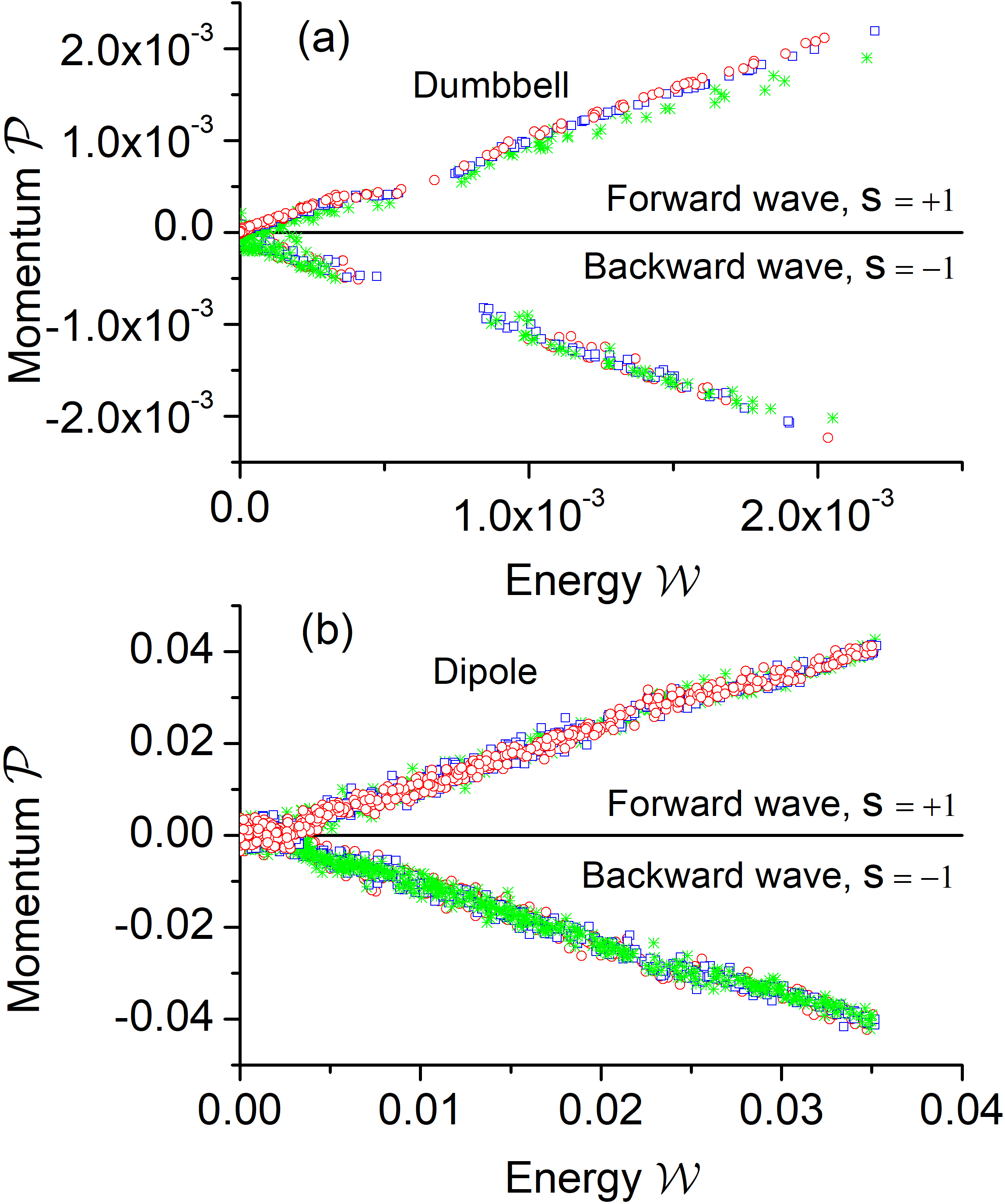}}
    \caption{Correlations between the energy and the momentum of composite dumbbell (a) and permanent-dipole (b) particles after the forward-wave ($s=1$) and backward-wave ($s=-1$) pulse passage. Red circles, blue squares, and green asterisks indicate results related to the parameter $\eta_g = 0.5$, 1.0, and 2.0, respectively.}
	\label{Energy-Pulse}
\end{figure}

To verify this result, we numerically solved the exact Eqs.~(\ref{cm2}) and (\ref{angle2}) for 500 composite particles, initially motionless and randomly oriented, $\varphi(0)\in(-\pi,\pi)$. We considered both forward ($s=1$) and backward ($s=-1$) wave packets with various group-velocity parameters $\eta_g$, as well as both dumbbell ($\sigma=1$) and permanent-dipole ($\sigma=-1$) particles. After the wave packet passes, the composite particle acquires final momentum $\mathcal{P}=2d\xi/d\tau$ and kinetic energy $\mathcal{W}=\mathcal{P}^2/4+I(d\varphi/d\tau)^2/2$, whose values are plotted in Fig.~\ref{Energy-Pulse}. One can see that these quantities satisfy the relation $\mathcal{P}/\mathcal{W}=s$ with a very good accuracy.   

\subsection{Induced dipole}

We now consider case (iii) of a \textit{polarizable} particle with induced dipole moment. The dynamics of such a particle in a 1D wave packet is described by a single translational degree of freedom $\xi$, which obeys Eq.~(\ref{cm2}) with $\sigma=-1$ and $d=\alpha \mathcal{F}$:
\begin{equation}
\label{eq_D8}
\frac{d^2\xi}{d\tau^2}=\alpha \mathcal{F}\frac{d\mathcal{F}}{d\xi}\,.
\end{equation}
This equation has quadratic right-hand side, and its straightforward averaging over fast wave oscillations yields 
\begin{equation}
\label{Fgrad}
\frac{d^2\bar{\xi}}{d\tau^2}=\frac{\alpha}{4}\frac{d a^2}{d\bar{\xi}} \equiv \mathcal{F}_{\rm pond}\,.
\end{equation}
This is a well-known gradient force that underpins optical or acoustic trapping of small particles \cite{Dholakia-2020}. It is conservative and has a form similar to that for a simple particle, Eqs.~\eqref{eq4} and \eqref{eq4a}. Once the wave packet passes, the initially motionless particle remains motionless, but its position is shifted towards or backwards the wave-packet source depending on the sign of the polarizability $\alpha$ and independently of the sign of the phase velocity, $s$.

\section{Beat waves}

In this Section, we examine potential applications of the ponderomotive forces in wavepacket-like fields to transport particles. Since a conservative gradient force induced by a single wavepacket shifts the simple particle, a suitable set of wave pulses can transport the particle over an arbitrary distance. 
Note that in contrast to a steady-force motion, which appears in monochromatic fields, here the particle experiences a stepwise motion.
It is particularly appealing to be able to control the transport direction using forward and backward wavepackets, similar to ``optical conveyors'' and ``tractor beams'' \cite{Chen-2011, Novitsky-2011, Ruffner-2012}.

Remarkably, particle manipulation by backward wave pulses has not been systematically explored so far.
Backward electromagnetic waves with anti-parallel phase and group velocities ($s=-1$) appear in the so-called double-negative (or left-handed) media \cite{Veselago-1967}, which are characterized by simultaneously negative permittivity $\varepsilon<0$  and permeability $\mu<0$, and are usually realized by complex periodic lattices composed of metal and dielectric elements \cite{Smith-2000}. 
There are also waveguides that support modes with negative phase velocity and have a free channel for electron beam transport \cite{Temkin-2016, Clarricoats-Waldron}. In addition, there are surface backward waves propagating along an interface between two media, such as plasma-vacuum interface \cite{Trivelpiece-1967}. 
In principle, particle manipulation can be realized in such structures, but the presence of unavoidable material elements makes this challenging.

\begin{figure}[t]
\centering{\includegraphics[width=\linewidth]{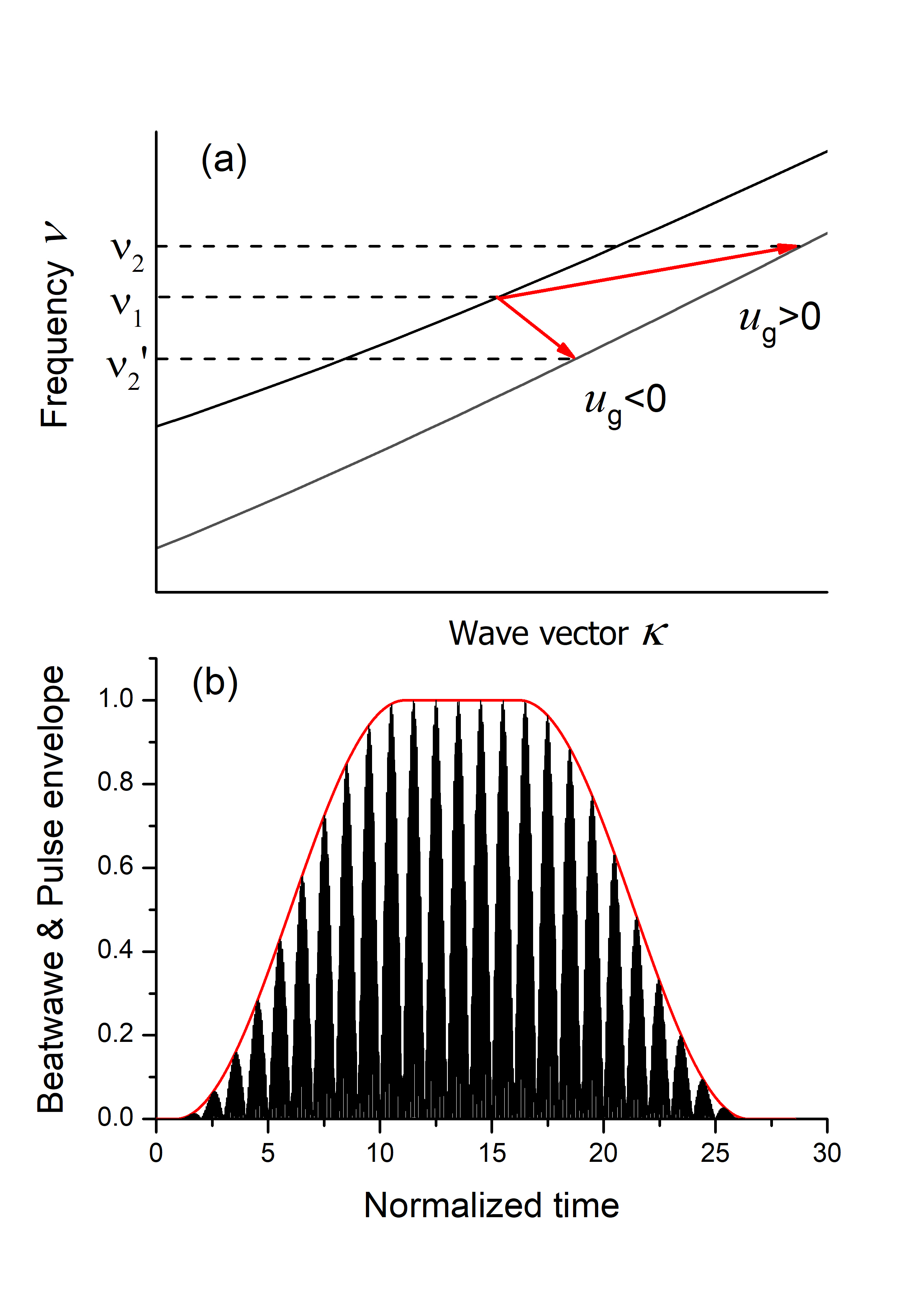}}
	\caption{(a) Schematics of the dispersion characteristics $\nu(\kappa)$ of two modes of a waveguide. Depending on the relation between the frequencies $\nu_1$ and $\nu_2$ ($\nu_2^\prime$), the beat wave has either positive or negative effective group velocity $u_g$. (b) The pulse envelope (red) and the beat wave sub-pulses (black). Time is normalized to the beat wave period $2\pi/\Delta\nu$.}
	\label{Fig2}
\end{figure}

Notably, there is another, much simpler method to control the effective direction of the wavepacket propagation, which mimics backward waves but does {\it not} require anti-parallel phase and group velocities. Namely, we consider a {\it beat wave}, which is produced by the interference of two modes with close frequencies and wavevectors:
\begin{equation}
\label{eq5}
\mathcal{F}(\xi,\tau)=a_1\cos(\kappa_1\xi-\nu_1\tau)-a_2\cos(\kappa_2\xi-\nu_2\tau).
\end{equation}
Here, the dimensionless frequencies $\nu_{1,2} = 1\pm \Delta \nu$ and wavevectors $\kappa_{1,2} = 1 \pm \Delta\kappa$ of the two waves obey different dispersion relations, such as, e.g., 
$\nu_{1,2}^2=\kappa_{1,2}^2+\nu_{c\,1,2}^2$ for two modes of a waveguide, with 
cut-off frequencies $\nu_{c\,1,2}$, see Fig.~\ref{Fig2}a.
The field \eqref{eq5} with $a_1=a_2\equiv a/2$, $|\Delta\nu| \ll 1$, $|\Delta\kappa | \ll 1$ can be written as:
\begin{equation}
\label{eq6}
	\mathcal{F}(\xi,\tau)=a\sin\left(\xi- \tau\right)\sin\left(\Delta\kappa\xi-\Delta\nu\tau\right),
\end{equation}
This field has a form of a traveling wave $\propto\sin\left(\xi-\tau\right)$ with periodic slowly varying amplitude $a\sin\left(\Delta\kappa\xi-\Delta\nu\tau\right)$, i.e., it mimics a \textit{periodic sequence of wavepackets} (hereafter, referred to as \textit{sub-pulses}). Here, the wave phase velocity $u_{\rm ph}=1$ is always positive, whereas the effective group velocity (i.e., the direction of motion) of sub-pulses $u_g=\Delta\nu/\Delta\kappa$ can be either positive or negative. (Note that the wave energy flow is always forward-directed.) Thus, one can control the motion of sub-pulses forward or backward to the wave source by varying the frequencies of two {forward-propagating waveguide modes} (Fig.~\ref{Fig2}a).

One can expect that the motions of a particle in a set of isolated wavepackets and in the beat wave (\ref{eq6}) are similar. There is a significant peculiarity though. 
In the isolated-wavepacket consideration, we assumed that the particle was at rest \textit{before} the pulse arrives. However, for an infinite set of periodic sub-pulses, the particle can either appear suddenly in the already existing wave (like an electron under ionization),  
or a gradually increasing front of the beat wave runs up at the initially motionless particle (i.e., the amplitude $a$ becomes a very slowly varying envelope). 

\begin{figure}[tbh]
	\centering {\includegraphics[width=\linewidth]{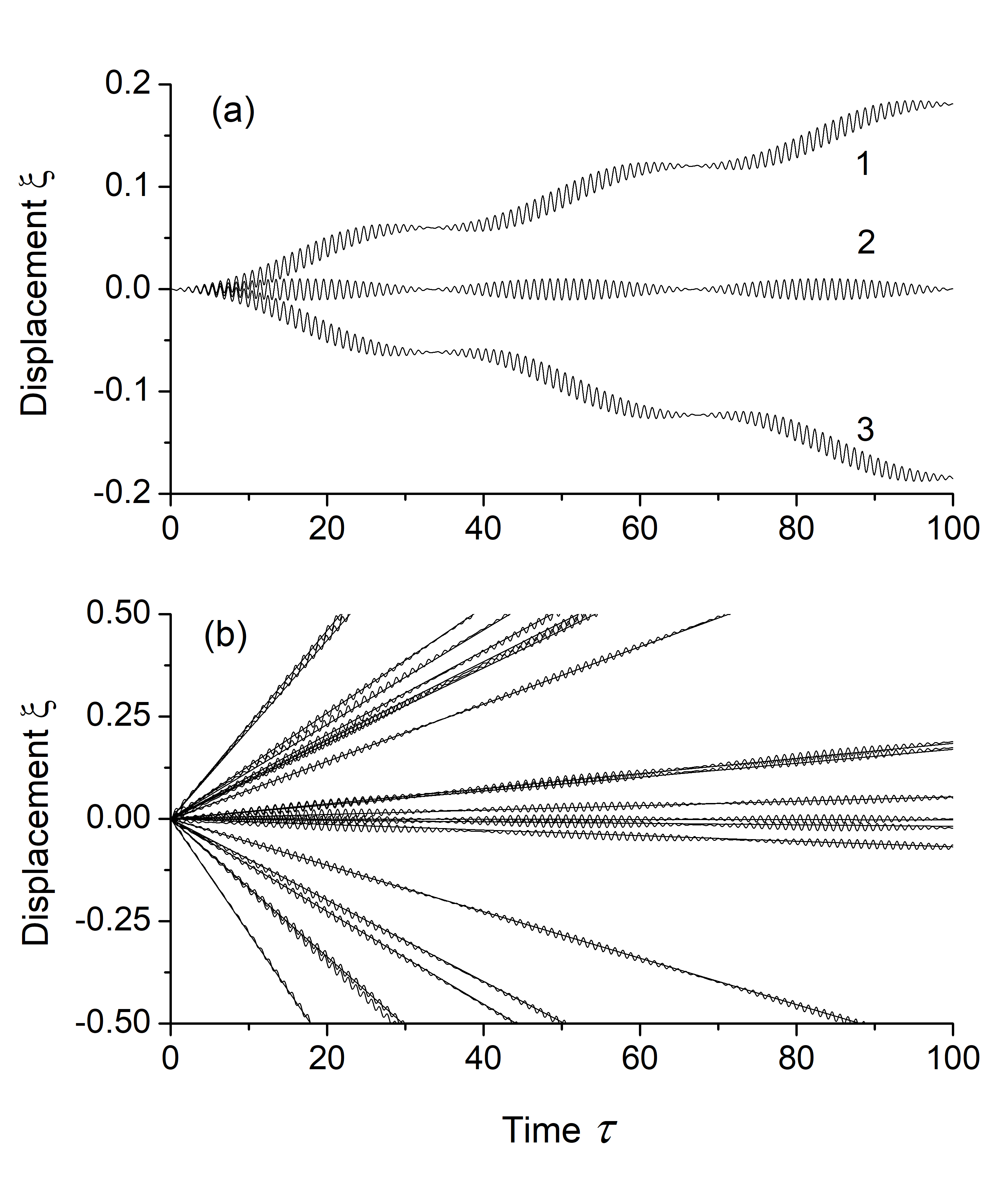}}
	\caption{Motion of simple point particles in a beat-wave with negative effective group velocity, $u_g<0$. (a) The particle is initially placed exactly in the beat-wave node. The curves 1, 2, and 3 correspond to the effective group velocity of the beat wave $|u_g|=0.55 > u_{\rm ph}/2$ (where $u_{\rm ph} =1$ is the phase velocity), $|u_g|=0.5 = u_{\rm ph}/2$, and $|u_g|=0.45 < u_{\rm ph}/2$, respectively. (b) The particles are initially randomly distributed along the wavelength near the beat-wave node. The value of the effective group velocity is not important in this case.}
	\label{Single-BW}
\end{figure}

In the first case, the particle motion strongly depends on the position of the particle's appearance. If the particle is placed in the beat-wave node, then every sub-pulse shifts the particle in the same manner as a single isolated wavepacket does. Figure~\ref{Single-BW}a demonstrates this with numerical solutions of the equation of motion \eqref{eq2} for a simple particle in the field \eqref{eq6}. However, if the particle is placed in an arbitrary position, it generally acquires a strong non-oscillatory velocity component, of the same order as the oscillatory one. The gradient force is weak and is unable to change the direction of this steady motion of the particle. Figure~\ref{Single-BW}b shows that the motion of such randomly placed particles depends only on the wave phase at the particle's appearance but not on the sub-pulse propagation direction.

\subsection{Simple particle}

We now consider the second case, where the beat wave amplitude is modulated as a `super-wavepacket' of a finite length, see Fig.~\ref{Fig2}b. The forward front of such a wavepacket moves in the positive direction independently of the effective group velocity of the sub-pulses, $u_g$. This front can also be considered as a source of the `super-gradient' force, different from the gradient forces at the sub-pulses. 

Let us examine how these two types of forces influence the motion of a simple particle.
The equation of motion has the same form as Eq.~(\ref{eq2}):
\begin{equation}\label{BW1}
	\frac{d^2\xi}{d\tau^2}=\mathcal{F}(\xi, \tau)\simeq A(\xi,\tau)\sin(\tau-\xi),
\end{equation}
where $A(\xi,\tau)=a(\xi-\eta_g\tau)\cos(\Delta\nu\tau-\Delta\kappa\xi)$ and the envelope $a(\xi,\tau)$ is a `super-slow' varying function. 
The time-averaged (over fast oscillations) equation of motions is obtained from Eq.~(\ref{eq4}):
\begin{align}
\label{BW2}
	\frac{d^2\bar{\xi}}{d\tau^2 } 
    & =-\frac{1}{2}A\left(\frac{\partial A}{\partial\xi}-2\frac{\partial A}{\partial\tau}\right)
	\nonumber\\
     & =-\frac{1}{4}(1+2\eta_g)\frac{da^2}{d\theta}\cos^2(\Delta\nu\tau-\Delta\kappa\bar{\xi})\nonumber\\
    &-\frac{1}{4}\Delta\kappa\left(1+2\frac{\Delta\nu}{\Delta\kappa}\right)a^2\sin(2\Delta\nu\tau-2\Delta\kappa\bar{\xi}).
\end{align}

Let the characteristic spatial scale $\mathcal{L}$ of the modulation $a(\xi - \eta_g \tau)$ satisfies {$$\Delta\kappa\mathcal{L}\gg1~~{\rm and}~~ \Delta\nu\mathcal{L}/\eta_g\gg1.$$} Then, one can perform the second time averaging over oscillations with parameters $\Delta\kappa$ and $\Delta\nu$. In doing so, the displacement $\bar{\xi}$ is separated into the new `fast' and `slow' components, $\bar{\xi}=\tilde{\bar{\xi}}+\bar{\bar{\xi}}$. 
Introducing variables 
$\tau^\prime=2|\Delta\nu|\tau$ and $\xi^\prime=2|\Delta\kappa|\bar{\xi}$, we rewrite Eq.~(\ref{BW2}) as follows:
\begin{align}
\label{BW3}
\frac{d^2\xi^\prime}{d{\tau^\prime}^2} &= -\frac{|\Delta\kappa|}{16\Delta\nu^2}(1+2\eta_g)\frac{da^2}{d\theta}\left[1+\cos(\tau^\prime-s'\xi^\prime)\right]\nonumber\\
&-\frac{s'}{8}\frac{\Delta\kappa^2}{\Delta\nu^2}\left(1+2\Delta\nu/\Delta\kappa\right)a^2\sin(\tau^\prime-s'\xi^\prime),	
\end{align} 
where $s'={\rm sgn}(u_g)$. The averaging of the first term on the right-hand side of Eq.~(\ref{BW3}) is obvious. The second term has the same form as the force $\mathcal{F}$ in Eq.~(\ref{eq2}), and its contribution in the averaged equation of motion is described by Eq.~(\ref{eq4}), where the amplitude $a$ is replaced by the coefficient at the sine function in  Eq.~(\ref{BW3}). 
As a result, in the original variables, the averaged equation Eq.~(\ref{BW3}) becomes:
\begin{align}\label{BW4}
	\frac{d^2\bar{\bar{\xi}}}{d\tau^2} &= -\frac{1}{8}\Biggl[(1+2\eta_g)
    \nonumber\\
    &+\frac{(1+2s'|u_g|)^2}{16u_g^2}(1+2s'\eta_g/|u_g|)a^2\Biggr]\frac{da^2}{d\theta},
\end{align}
where $\theta=\bar{\bar{\xi}}-\eta_g\tau$.
 
Since the particle velocity is small, one can approximate $\theta\simeq -\eta_g\tau$ and $d/d\theta \simeq -\eta_g^{-1}d/d\tau$ along the particle trajectory. Then, the particle's average velocity $\bar{\bar{u}}_p=d\bar{\bar{\xi}}/d\tau$ can be found by the integration of Eq.~(\ref{BW4}) with respect to $\tau$:
\begin{equation}\label{BW6}
	\bar{\bar{u}}_p=\frac{1}{8\eta_g}\!\left[(1+2\eta_g)+\frac{(1+2s'u_g)^2}{32u_g^2}\!\left(1+2s'\frac{\eta_g}{u_g}\right)\!a^2\right]\!a^2.
\end{equation}
Expression (\ref{BW6}) shows that this velocity is always positive when $s'=1$, and can be negative when $s'=-1$ and the wavepacket amplitude exceeds the critical value given by
\begin{equation}\label{BW7}
	a_{\rm crit}=4 |u_g|\sqrt{\frac{2 u_g(1+2\eta_g)}{(1-2u_g)^2(2\eta_g-u_g)}}\,.
\end{equation}
This means that the modulated beat wave can provide controlled forward and backward transport of a simple particle. Within this simple general conclusion, the behavior of the particle exhibits a number of peculiarities.  

\begin{figure}[t]
	\centering {\includegraphics[width=\linewidth]{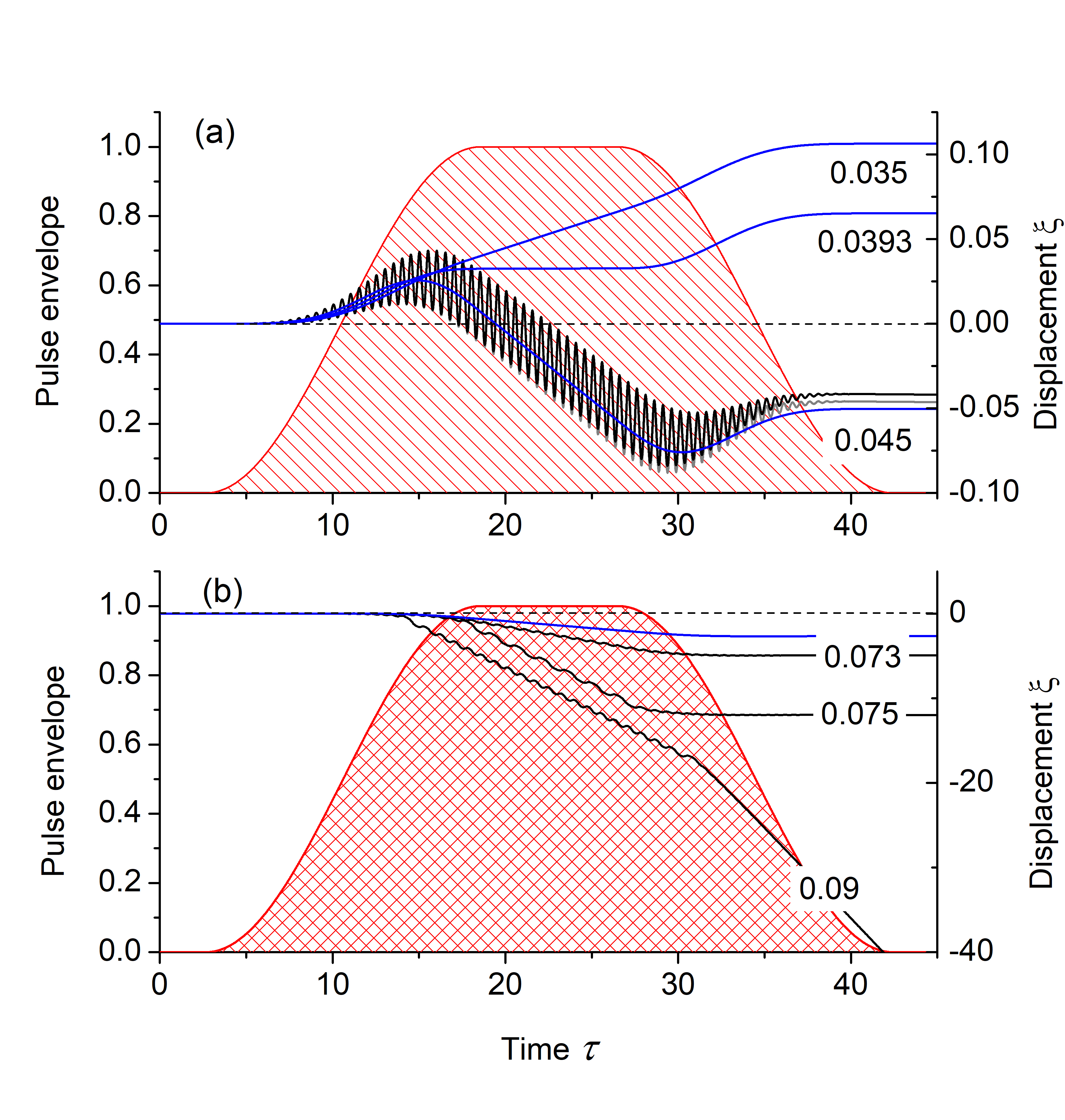}}
	\caption{Trajectories of a simple particle in beat wave packets with $\eta_g=0.9$, $|\Delta\nu|=3\cdot10^{-4}$, $|\Delta\kappa|=10^{-2}$, negative group velocity of sub-pulses, $s'=-1$, and different amplitudes $a_0$ (shown near the corresponding curves). 
     The pulse envelope is shown by the red hatched area.
     (a) $a_0=0.035 <  a_{\rm crit}$,  $a_0=0.0393 = a_{\rm crit}$ (the particle velocity vanishes inside the pulse), and $a_0=0.045 > a_{\rm crit}$. Solutions of exact Eq.~(\ref{BW1}), and averaged Eq.~(\ref{BW2}) are shown by practically indistinguishable oscillating gray and black curves (only for $a_0=0.045$). Solution of the twice-averaged Eq.~(\ref{BW4}) is shown by a smooth blue curve. (b) The same for larger pulse amplitudes: $a_0=0.73 > a_{\rm crit}$, but the particle is not trapped (the particle velocity is close to the group velocity of sub-pulses, and Eq.~(\ref{BW4}) loses its validity); $a_0=0.75$ and $a_0=0.09$ -- the particle is trapped by the ponderomotive potential well, so that its velocity inside the pulse is equal to the sub-pulses group velocity $u_g$ and is independent of the pulse amplitude.}
	\label{Single-BW-2}
\end{figure}

Figure \ref{Single-BW-2}a shows the motion of a simple particle in the backward propagating ($s'=-1$) beat wave packets with different amplitudes. One can see that solutions of the original equation of motion (\ref{BW1}) (black), the averaged equation (\ref{BW2}) (gray), and the twice-averaged equation (\ref{BW4}) perfectly agree with each other. Note the increase of particle's velocity at the leading and trailing edges of the wavepacket when $a < a_{\rm crit}$. The particle remains motionless ($\bar{\bar{u}}_p=0$) inside the wavepacket when $a = a_{\rm crit}$.  

\begin{figure}[t]
	\centering {\includegraphics[width=\linewidth]{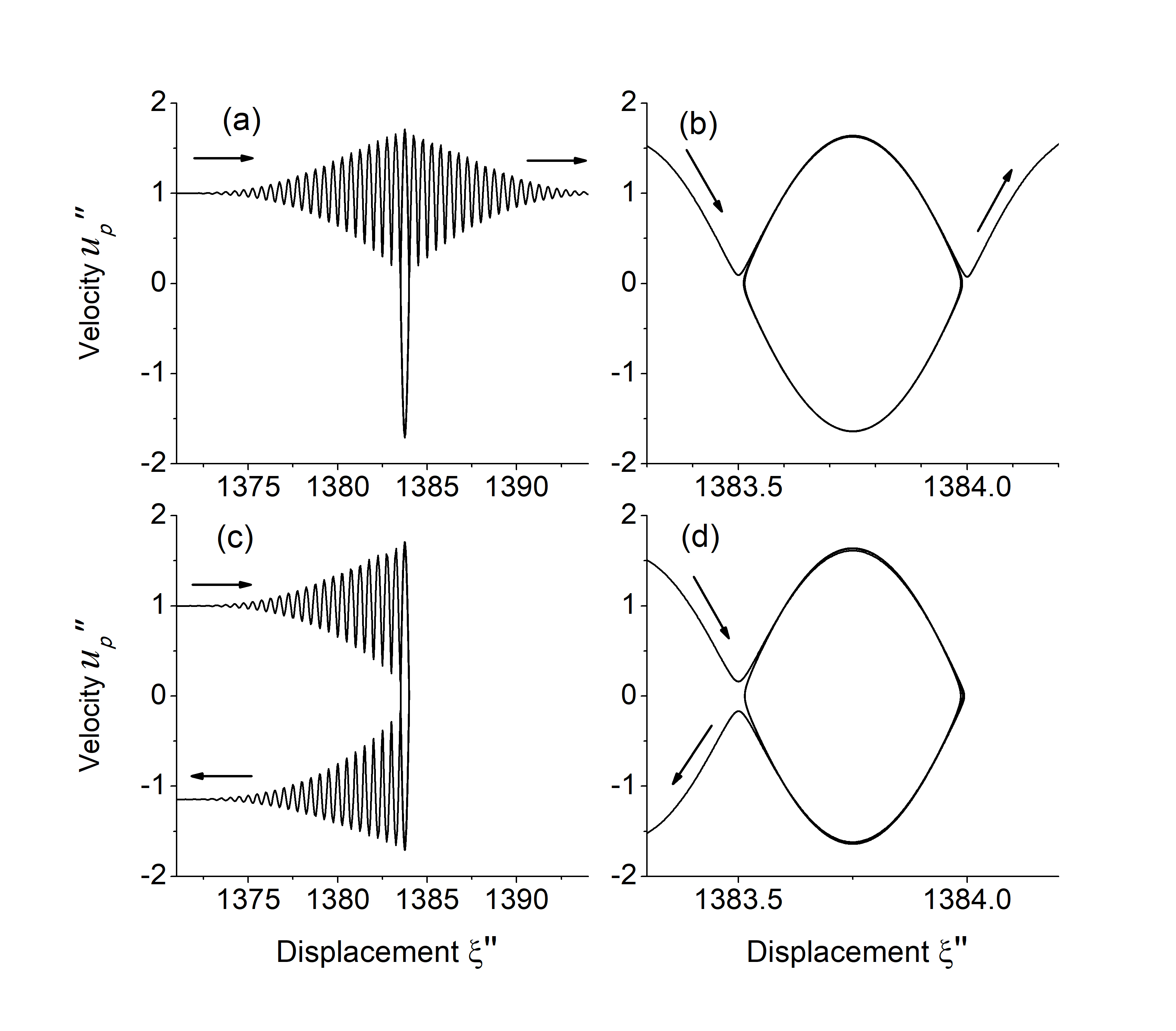}}
	\caption{Phase-space trajectories of a simple particle in the reference frame of the moving potential well. The particle is trapped by the potential well and leaves the well at the trailing edge of the pulse. (a) The particle leaves the potential well over the right barrier; (b) zoom-in view of the central part of (a) with the trajectory of the trapped particle. (c) and (d) The same as in (a) and (b) but the trapped particle leaves the potential well over the left barrier. }
	\label{Phase space}
\end{figure}

If the amplitude $a$ is so large that the particle's velocity is close to the sub-pulses group velocity $\bar{\bar{u}}_p\simeq u_g$, the particle is trapped by the ponderomotive potential well between the sub-pulses. Then, it oscillates in this well, and moves with the well velocity $u_g$, whose value is independent of $a$, as shown in Fig.~\ref{Single-BW-2}b. 
In the reference frame of the moving potential well, the potential magnitude is a slowly varying function of time. In the beginning, when the potential magnitude is small, the particle's motion is unbounded. When the potential magnitude becomes large enough, this motion is changed into localized oscillations in the potential well. In the rear end of the wavepacket, the particle is released from the potential well. Due to the adiabaticity of this process, the energy of the particle (in the potential-well frame) remains the same as it was before the wavepacket arrival. However, the particle can leave the potential well over its right or left barrier (in a random fashion). Correspondingly, the final particle's velocity can be equal to either $u_g$ or $-u_g$, and in the laboratory reference frame, it becomes either ${u}_{p\, {\rm fin}}=0$ or ${u}_{p\, {\rm fin}}=-2u_g$, as can be seen in Fig.~\ref{Phase space}.

To determine the amplitude $a$ sufficient to trap the particle, let us write Eq.~(\ref{BW3}) in the reference frame of the ponderomotive potential well, using the substitution $\xi^\prime=\xi^{\prime\prime}+s'\tau^\prime$ and  $\theta=\xi-\eta_g\tau = \left[\xi^{\prime\prime}-\left({\eta_g}/{u_g}-s'\right)\right]/{2|\Delta\kappa|}$:
\begin{align}\label{Tr1}
	\frac{d^2\xi^{\prime\prime}}{d{\tau^\prime}^2} &=-\frac{1}{8u_g^2}\bigg[(1+2\eta_g)\frac{\partial a^2}{\partial\xi^{\prime\prime}}(1+\cos\xi^{\prime\prime})\nonumber\\
    &+(1+2s'|u_g|)a^2\sin\xi^{\prime\prime}\bigg].	
\end{align} 	

The two terms with sine and cosine functions in Eq.~\eqref{Tr1} can be combined in a single term $\propto\sin(\xi^{\prime\prime}+\phi)$ with slowly varying phase $\phi\sim a^{-2}(\partial a^2/\partial\xi^{\prime\prime})\ll 1$. Neglecting this phase $\phi$, i.e., neglecting the term with cosine, Eq.~(\ref{Tr1}) can be written as:
\begin{equation}\label{Tr2}
\frac{d^2\xi^{\prime\prime}}{d{\tau^\prime}^2}=-\frac{\partial \mathcal{U}}{\partial\xi^{\prime\prime}},
\end{equation}	
where
\begin{equation}\label{Tr3}
	\mathcal{U}=\frac{a^2(\tau^\prime)}{8u_g^2}\left[(1+2\eta_g)+(1+2s'|u_g|)\cos\xi^{\prime\prime}\right]
\end{equation}
is the ponderomotive potential of the beat wave. If the amplitude $a$ is constant, then Eqs.~(\ref{Tr2}) and (\ref{Tr3}) can be integrated:
\begin{equation}\label{Tr4}
	{u_p^{\prime\prime}}=\pm\sqrt{2[\mathcal{W}-\mathcal{U}(\xi^{\prime\prime})]},
\end{equation}
where  $u^{\prime\prime}_p=d\xi^{\prime\prime}/d\tau^\prime$ is the particle velocity, and $\mathcal{W}$ is the constant total energy of the particle. Equation (\ref{Tr4}) describes the  particle trajectory $u_p^{\prime\prime}(\xi^{\prime\prime})$ in the phase space $(\xi^{\prime\prime}, u^{\prime\prime}_p)$, even when the amplitude $a$ varies slowly with time. {However, the  energy $\mathcal{W}$ is not constant now. }To define the corresponding variation of the energy $\mathcal{W}(\tau^\prime)$, one can use the adiabatic invariant (the action) {of the motion in a spatially-periodic potential} $\mathcal{I}=\intop_0^{2\pi}u_p^{\prime\prime}d\xi^{\prime\prime}$, whose value is defined by the initial state of the particle. If the particle was motionless before the pulse arrived, then $u^{\prime\prime}_p (0)=-s'$ in the chosen variables, $\mathcal{W}(0)=1/2$, $\mathcal{U}(0)=0$, and the initial action for a freely moving particle is $\mathcal{I}=-2s'\pi$. One the other hand, the action can be calculated via the integral
\begin{align}
\label{Tr5}
&\mathcal{I} =-2s\pi \nonumber\\ 
&=\!-s'\!\intop_0^{2\pi}d\xi^{\prime\prime}\!\sqrt{2\mathcal{W}-\frac{(1+2\eta_g)a^2}{4u_g^2}
-\frac{(1+2s'|u_g|)a^2}{4u_g^2} \cos\xi^{\prime\prime}} \nonumber \\
 &=4\sqrt{2(\mathcal{W}^\prime+\mathcal{U}^\prime)}\,{\rm \bf E}\!\left(\sqrt{2\mathcal{U}^\prime/(\mathcal{W}^\prime+\mathcal{U}^\prime)}\right)\!,
\end{align}
where {\rm \bf E} is the elliptic integral of the second kind, $\mathcal{W}^\prime=\mathcal{W}-(1+2\eta_g)a^2/8u_g^2$, and $\mathcal{U}^\prime=(1+2s'|u_g|)a^2/8u_g^2$. {As the amplitude $a$ increases with time, the current value of the energy, $\mathcal{W}^\prime$, decreases, while the height $\mathcal{U}^\prime$ of the potential barrier increases. When $\mathcal{W}^\prime=\mathcal{U}^\prime$, the initially transit particle becomes trapped, and its trajectory reaches the separatrix.} 
Using Eq.~(\ref{Tr5}), and taking into account that ${\rm \bf E}(1)=1$, this condition can be written as $\sqrt{\mathcal{U}^\prime}=\pi/4$, which yields
\begin{equation}\label{Tr8}
a_{\rm trap}= |u_g|\frac{\pi}{\sqrt{2(1+2s'|u_g|)}}.
\end{equation}	
Numerical solution of Eq.~(\ref{BW1}), presented in Fig.~\ref{Single-BW-2}b, shows that the particle is trapped when the wavepacket amplitude $a > 0.074$. This result agrees well with Eq.~(\ref{Tr8}) producing the value $a_{\rm trap}\simeq 0.069$ for the parameters used in Fig.~\ref{Single-BW-2}.

\subsection{Polarizable particle}

The motions of simple and polarizable particles differ little from each other. Indeed, Eq.~(\ref{eq_D8}) and Eq.~(\ref{BW2}) have similar structures, except for opposite signs in the right-hand side and the absence of the term $\propto\partial A/\partial\tau$ in Eq.~(\ref{eq_D8}). As a result, the time-averaged equation of motion for the polarizable particle takes the form:
\begin{align}
\label{BW10}
\frac{d^2\bar{\xi}}{d\tau^2} &=
\frac{1}{2}\overline{A\frac{\partial A}{\partial\xi} }
	=\frac{1}{8}\frac{da^2}{d\theta}\biggl[1+\cos(2\Delta\nu\tau-2\Delta\kappa\bar{\xi})\nonumber\\
   & +\frac{1}{4}\Delta\kappa a^2\sin(2\Delta\nu\tau-2\Delta\kappa\bar{\xi})\biggr].	
\end{align}
Here, the polarizability  coefficient $\alpha$ is eliminated by the amplitude re-normalization $a^2 \to \alpha a^2$. 

\begin{figure}[t]
\centering {\includegraphics[width=\linewidth]{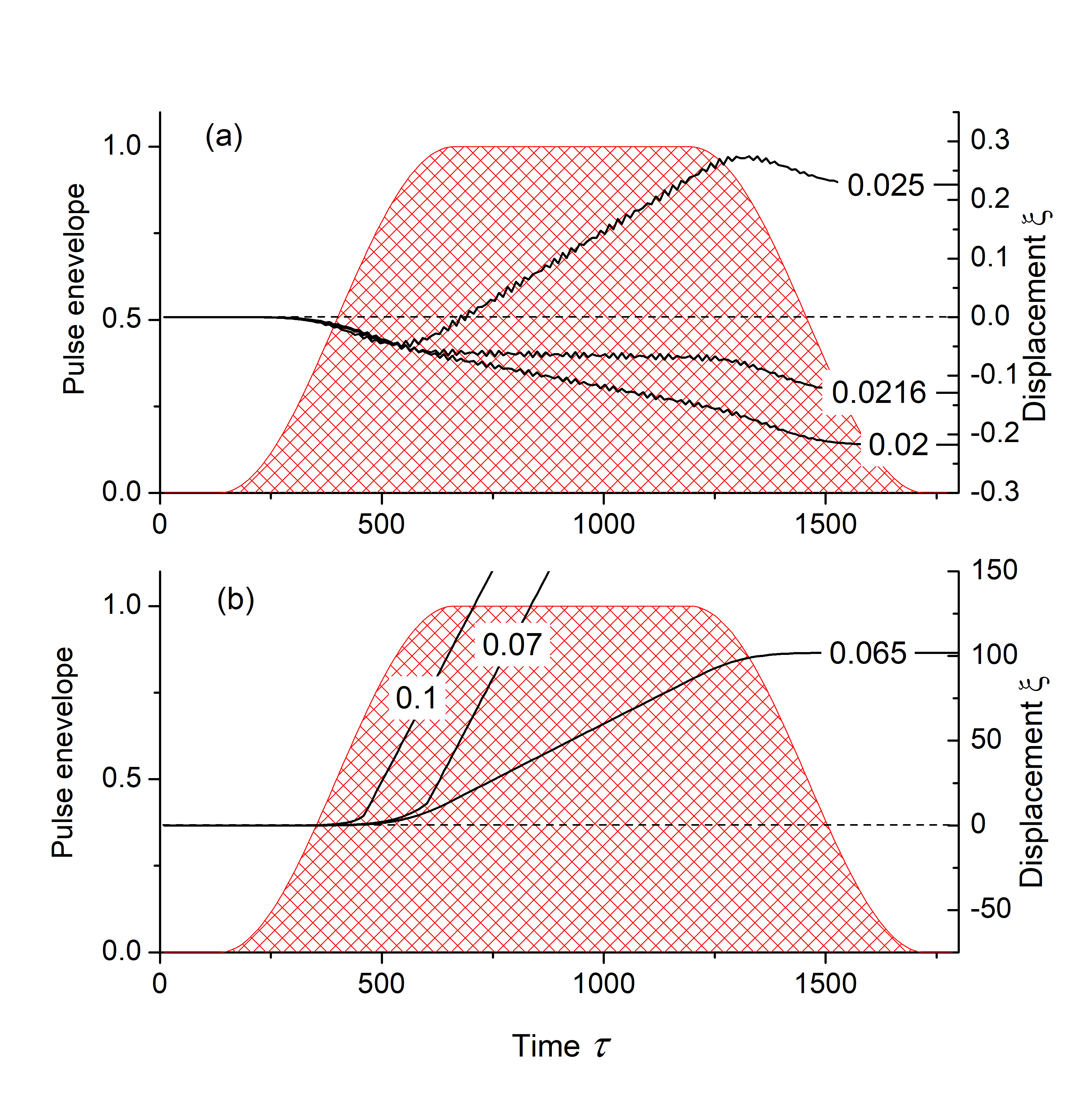}}
\caption{Numerical solutions of Eq.~\eqref{eq_D8} providing exact trajectories of a polarizable particle in the beat wave pulses with different amplitudes $a_0$ (shown near the corresponding curves) and positive group velocity of sub-pulses ($s'=+1$). The wavepacket parameters and notations are the same as in Fig.~\ref{Single-BW-2}, which corresponds to $a_{\rm crit}\simeq 0.0216$ and $a_{\rm trap}\simeq 0.067$ in Eqs.~\eqref{BW10a}.  
}
	\label{Polarizable-BW}
\end{figure}

Note that the replacements $(1+2\eta_g)\to -1$ and $(1+2s' |u_g |)\to -1$ transform Eq.~(\ref{BW4}) into Eq.~(\ref{BW10}). Therefore, the motion of a polarizable particle in a {\it forward}-propagating beat wave $(s'=+1)$ is entirely similar to the motion of a simple particle in the corresponding {\it backward}-propagating beat wave $(s'=-1)$. This determines three regimes of the particle dynamics: negative shift after the super-wavepacket passage for $a<a_{\rm crit}$, positive shift for $a>a_{\rm crit}$, and trapping (with a positive shift) for $a>a_{\rm trap}>{a_{\rm crit}}$, where
\begin{align}
\label{BW10a}
a_{\rm crit}=4 u_g\sqrt{\frac{2 u_g}{u_g+2\eta_g}}\,,\quad 
a_{\rm trap}=u_g\frac{\pi}{\sqrt{2}}\,.
\end{align} 
Figure~\ref{Polarizable-BW} shows numerical calculations of a polarizable particle motion in a forward-propagating super-wavepacket of a beat wave, which agrees well with analytical predictions. 
For the beat wave pulse with the parameters specified above $a_{\rm crit}=0.0216$, and  $a_{\rm tr}=0.067$, that agrees well with the results of numerical solutions of Eq.~(\ref{BW10}), presented in Fig.~\ref{Polarizable-BW}.

\subsection{Particles sorting}

The motion of a simple particle in a backward-propagating beat wave, and the motion of a polarizable particle in a forward-propagating beat wave, demonstrate an unusual property: the direction of the particle shift can be easily controlled by the wave amplitude, see Figs.~\ref{Single-BW-2}a and \ref{Polarizable-BW}a. This property can be used for particle \textit{sorting}, i.e., the separation of particles with different physical characteristics \cite{Brzobohaty-2013, Sorting-2025, Toftul-2024}.

Since equations in this paper deal with the acceleration $d^2\xi/d\tau^2$, the field amplitude $a$ is actually inversly proportional to the particle mass $m$. This means that amplitude-dependent behavior enables effective sorting by the particle density or size. Furthermore, the effective polarizability $\alpha$ of small optical or acoustic particles is also proportional to the particle's volume and depends on the material characteristics \cite{Toftul-2024, Kandemir-2021}.

\subsection{Comparison with previous works} 

Beat waves and analogous setups have been discussed and employed for manipulation of particles in different contexts. 
In particular, beat waves can be formed by the interference of two non-collinear plane waves or wave beams with slightly different frequencies. For example, a scheme using two non-collinear laser beams for the acceleration of electrons has been suggested in \cite{Sakai-1989, Sakai-1991} (see also \cite{Hafizi-1997}). However, these works considered only the acceleration of electrons; a backward-moving beat wave and pulling forces have not been studied.   

The motion of dielectric particles in the field of two laser beams with \textit{equal} frequencies but different longitudinal wavevector components was considered in \cite{Cizmar-2006, Ruffner-2012, Brzobohaty-2013}. Such a field generates stationary spatially-periodic potential wells. Then, the phase shift between the beams can move the particles captured in the wells. The continuous motion of the trapped particle can be achieved by a continuous phase shift, i.e., a frequency shift \cite{Sadrove-2016}. 
{A 2D generalization of these ideas was recently demonstrated for floating particles in gravity-capillary water waves \cite{Wang2026A}. It should be noticed that these earlier works only considered the motion of particles trapped in potential wells. In the present work, we examine a more general case, where the particle is not necessarily trapped.} 

{In addition, polyphonic acoustic fields can create dynamic acoustic traps, so-called ``spectral holography'' \cite{Abdelaziz-2021, Morrell-2024}. Two or more sound sources with slightly different frequencies and different wavevectors form a mobile potential landscape enabling transport of the trapped particles.} 

\section{Conclusions}  

We have described the dynamics of particles of various types in fast-oscillating fields of 1D quasi-monochromatic wavepackets and beat-wave packets. The ponderomotive force (which is sometimes referred to as the gradient force) depends to the same extent on slow spatial and temporal variations of the wave amplitude. 
Our simplified models can be applied to charged particles (`simple particle'), anisotropic-shape particles (`dumbbell'), permanent-dipole particles, as well as polarizable particles (`induced dipole'). We paid particular attention to backward waves with oppositely-directed phase and group velocities, and to beat waves, in which the direction of the motion of sub-pulses can be controlled by the wave frequencies. {The beat wave, formed, e.g., by two different forward-propagating eigenmodes of a waveguide, exhibits properties of a backward-propagating wave. (Experimental demonstration of the particle pulling based on such a backward wave is an open challenge.)}
Despite the simplicity of the 1D system under consideration, it exhibits a very rich behavior.  
We have provided a thorough theoretical analysis of various regimes of the particle dynamics, as well as numerical solutions of the equations of motion.

In particular, the motion of a simple point particle in a wavepacket field is conservative and depends on the forward or backward phase velocity of the wave. 
Next, dumbbell and permanent-dipole particles can demonstrate either regular or chaotic regimes depending on the wave amplitude. 
Finally, the dynamics of various particles in a beat wave crucially depends on the wave amplitude and provides trapping in a moving potential well for high amplitudes. 

Importantly, our models describe different mechanisms for the control of the particle motion from the wave source (pushing) or towards the source (pulling). In particular, small variations in the frequencies or amplitudes of interfering waves can readily alter the direction of the particle motion. We have found that the process of particle insertion in the beat wave field significantly affects its motion, but this difficulty can be overcome by using a `super-wavepacket' with adiabatic modulation of the beat-wave amplitude. 

{In the present work, we considered the main universal models, which can be applied to various wave systems interacting with small particles: optical, acoustic, and even to floating particles in water waves \cite{Falkovich2005, Wang2025N, Wang2026A}. In optics, the induced-dipole forces are primary for isotropic particles, whereas in acoustics, the monopole contribution (caused by the field-induced compression-extension of the particle) is also important \cite{Toftul-2024}. In the framework of our paper, such acoustic monopole force can be modeled by an induced dumbbell (without rotational degree of freedom), whose dynamics is similar to the induced-dipole case.}

\section*{Acknowledgments} I am grateful to my colleague and son, Konstantin Bliokh, for his critical comments, which greatly improved the quality of the paper.

\section*{DATA AVAILABILITY}

The data are available upon reasonable request from the
authors.

\bibliography{Bib-final}

\end{document}